\documentstyle[preprint,prb,aps]{revtex}

\begin{document} 
\draft
\title{Correlations, compressibility, and capacitance in
double-quantum-well systems in the quantum Hall regime}

\author{T. Jungwirth}
\address{Institute of Physics ASCR,
Cukrovarnick\'{a} 10, 162~00 Praha 6, Czech Republic\\}
\author{A.H. MacDonald}
\address{Department of Physics, Indiana University,
Bloomington, IN 47405}
\date{Received \today}

\maketitle

\begin{abstract}

In the quantum Hall regime, electronic correlations in double-layer 
two-dimensional electron systems are strong because the 
kinetic energy is quenched by Landau quantization.  In this
article we point out that these correlations are reflected in  
the way the partitioning of charge between the two-layers
responds to a bias potential.
We report on illustrative calculations based on an unrestricted 
Hartree-Fock approximation which allows for  
spontaneous inter-layer phase coherence.  The possibility of 
studying inter-layer correlations by capacitive coupling to 
separately contacted two-dimensional layers is discussed
in detail. 

\end{abstract} \pacs{}

\section{Introduction}

Technological progress has made it possible 
to fabricate epitaxially grown semiconductor systems
with nearby two-dimensional electron layers
and has led to interest in the physics of the various inter-layer 
coupling effects which occur as a consequence.  
As shown in Figure~\ref{fdw}, these systems consists of two parallel
electron layers confined by narrow rectangular quantum wells.  In
standard GaAs/AlGaAs structures with the width of the wells of order
10 nm and the barrier height about 250 meV, electron wavefunctions are
strongly localized around the center of each quantum well and the
overlap between layers is very small.  To date coupling effects have 
been observed primarily in the transport properties of double-layer
systems.  For example, inter-layer electron-electron
interactions lead\cite{dragexp,dragth} to frictional drag voltages when charge
in one layer is moved relative to charge in the nearby layer.
Inter-layer tunneling leads to quantum interference effects which 
are responsible for interesting dependence of both in-plane\cite{inplane}
and inter-plane\cite{interplane} conductances on the strength
of a magnetic field oriented parallel to electron layers.
In a strong perpendicular magnetic field, the kinetic energy
of the electrons is quenched by Landau quantization and, 
at least in high-mobility systems, electron-electron interactions dominate the 
physics.  For double-layer systems inter-layer interactions are 
responsible for novel broken symmetries\cite{dlprl} and, if tunneling
between layers also occurs, for inordinate sensitivity to small tilts
of the field away from the normal to the electron layers.\cite{murphy}

In this paper we discuss the effect of inter-layer coupling 
on {\em equilibrium} properties of double-layer systems.  In particular 
we consider the variation of the partitioning of charge 
between the two-layers as the total electron density is 
modified by adjusting an external gate potential.  
Eisenstein and co-workers\cite{jimcompress} have measured this 
quantity for the case of more remotely spaced layers by combining a
standard capacitive method with a measurement of the charge 
transferred between layers when the gate voltage is changed.
Using the assumption (valid
in that work) that inter-layer correlations could safely be neglected, 
they were able to relate the measured inter-layer current to the 
compressibility of the electron layer closest to the gate.
In Section II we use an idealized model 
with infinitely narrow quantum wells to generalize their analysis  
to the case where inter-layer correlations are important.  In a strong 
perpendicular magnetic field, the electronic properties 
of double-layer systems are extremely subtle.  To date 
most studies\cite{dlnum,amdreview} of double-layer systems have focused on
systems with equal density in each layer.  In this article we use
an unrestricted Hartree-Fock approximation to obtain qualitative results as 
a function of layer separation over the 
full range of total filling factors and bias potentials in the quantum
Hall regime.  The Hartree-Fock approximation allows for 
spontaneous inter-layer phase coherence\cite{dlprl} and is developed from
two different points of view in Sections~III and IV.  
In Section~V we present and discuss the results predicted for Eisenstein's
double-layer capacitance measurement by the unrestricted Hartree-Fock
approximation.  Finally, we present our conclusions in Section~VI.

\section{Narrow-Well Double-Layer Model}

In this section we assume that only the lowest 
energy subband is relevant in each quantum well and, for 
convenience, we take the two quantum wells to be identical.
We further assume that each quantum well is sufficiently narrow
that we can replace the charge density in each 
by a zero-thickness layer located at the center of 
the quantum well.  With these assumptions it follows that for 
fixed external charges (assumed to reside away from the 
double layer system) the energy of the double-layer system is 
given up to an irrelevant constant by 
\begin{equation}
{E \over A} = \frac{e^2 d}{2\epsilon}
 (N_R - N_0)^2 + \varepsilon(N-N_R,N_R),
\label{eq:amd1}
\end{equation}
where $A$ is the area of the system, $N_R$ and $N_L= N - N_R $
are the areal densities of 
electrons in the right and left layers, $N$ is the total electron 
density and $N_0$ is determined by external charges as
discussed below.  In Eq.~(\ref{eq:amd1}) $\varepsilon(N_L,N_R)$ would be
the energy  per area of the double-layer system if neutralizing external
charges were located in each layer of the double-layer system.
This quantity is the conventional point of contact between electron gas
theory and experiment.  For a given configuration of external charge,
the charge distribution is determined by minimizing the sum of 
$\varepsilon(N_L,N_R)$ and the electrostatic energy.
The zero-thickness layers, with areal charge densities 
$eN_L$ and $eN_R$, yield discontinuities
in the dependence of the electric field along the direction between layers
(which we take to be the $\hat z$ direction) across each layer.
We assume that any charges induced by variation of the gate voltage go entirely
into the electron layers so the electric field $E_0$ 
at the right boundary of the double-layer in Figure~\ref{fdw} is 
independent on the voltage and enters the problem as an input
parameter.  From the Poisson equation we then obtain
\begin{eqnarray}
E_1 &=& E_0-\frac{|e|}{\epsilon}N_R \equiv \frac{|e|}{\epsilon}(N_0 - N_R) 
\nonumber \\
 E_2 &=& E_0- \frac{|e|}{\epsilon}(N_R+N_L) .
 \label{a}
\end{eqnarray}
$N_0$ is defined by this equation. 
Note that changing $E_2$ is equivalent to changing $N = N_R + N_L$. 
(See Figure~\ref{fdw}.) 

The double-layer capacitance technique of Eisenstein {\em et al.}
measures $R_E$,  the ratio of the electric field change between
the electron layers to the electric field change 
between the gate and the nearest electron layer: 
\begin{equation}
R_E \equiv {dE_1 \over dE_2} = {dN_R \over dN}.
\label{eq:amd2}
\end{equation}
Given $N$, \, $N_R$ is determined by minimizing the total energy
described in Eq.~(\ref{eq:amd1}) yielding,
\begin{equation} 
\mu_L(N-N_R,N_R)= \mu_R(N-N_R,N_R) + \frac{e^2 d }{\epsilon}  
(N_R - N_0),
\label{eq:amd3}
\end{equation}  
where 
\begin{equation}
\mu_L(N_L,N_R)  \equiv \frac{\partial \varepsilon(N_L,N_R)}{\partial N_L}
\label{eq:amd4}
\end{equation}
and $\mu_R(N_L,N_R)$ is defined similarly.  $\mu_L(N_L,N_R)$ includes
all contributions to the chemical potential for electrons in the left
layer except for the contribution from the electrostatic potentials and 
would be the full chemical potential if, as in conventional
electron gas literature, neutralizing positive charges in
each layer were assumed.  It follows from Eq.~(\ref{eq:amd3}) that 
\begin{equation}
R_E = \frac{ d_{LL} - d_{RL}}{ d + d_{LL} + d_{RR} - d_{RL} - d_{LR} } 
\label{eq:amd5}
\end{equation}
where we have followed Eisenstein {\em et al.}
introducing\cite{jimcompress} a set of lengths defined by 
\begin{equation}
d_{AB}(N_L,N_R)  \equiv \frac{\epsilon}{e^2} 
\frac{\partial \mu_A(N_L,N_R)}{\partial N_B}.
\label{eq:amd6}
\end{equation}
In Eq.~(\ref{eq:amd6}) $A$ and $B$ are layer labels.  When inter-layer 
electron-electron interactions can be neglected $d_{LR}=d_{RL}=0$ and 
\begin{equation}
d_{AA}(N_A)  = \frac{\epsilon}{e^2 \kappa_A N_A^2 } 
\label{eq:amd7}
\end{equation}
where $\kappa_A$ is the compressibility of the electron system in 
layer $A$ with the usual convention of a neutralizing background.
For non-interaction electrons and zero magnetic field
$ d_{AA} = d_E \equiv a_0^{*}/4 $
is independent of the electron density in layer $A$; here $d_E$ is
the length defined by Eisenstein {\em et al.} and 
$a_0^{*} = \hbar^2 \epsilon/ m^{*} e^2$ is the effective Bohr radius
of the semiconductor.  For $GaAs$ $a_0^{*} \approx 10 $ nm so 
$d_E \approx 2.5 $ nm.  For non-interacting electrons in a 
strong magnetic field, $ d_{AA} = 0$ when a Landau level is 
partially filled and $ d_{AA} = \infty$ at integer Landau level
filling factor.

\section{Unrestricted Hartree-Fock Approximation for Interlayer
Correlations: Self-Consistent Field Equation Approach} 

For decoupled layers, electron-electron
interactions can reduce or even\cite{jimcompress} change the sign of $d_{AA}$.
In the following sections we discuss the effect of inter-layer coupling 
on $R_E$.   In the absence of a magnetic field inter-layer interactions
have little effect\cite{lian} on $R_E$ at experimentally accessible
layer separations.  The situation is different at strong magnetic fields
where the kinetic energy of the electrons is quenched and interaction 
effects are very strong.  The problem of finding accurate results
for the dependence of the ground state energy in this regime 
on the density in each layer and on the layer separation 
is a difficult one which is largely unsolved because perturbative 
approaches are unsuitable.  Numerical exact-diagonalization results
can provide guidance and some results\cite{dlnum} are already 
available using this approach.  In the following sections we follow
an alternate line by developing a simple Hartree-Fock approximation for 
biased double-layer systems.  In our Hartree-Fock approximation inter-layer
correlations can be generated by forming broken-symmetry states
with spontaneous inter-layer phase coherence, as we describe in more
detail below.  Such a broken symmetry does in fact\cite{dlprl} occur in 
double-layer systems in strong magnetic fields,
although not over as wide a range of densities and layer 
separations as in our calculations.  The correlations which
appear only in connection with a broken symmetries in the single
Slater determinant states of the Hartree-Fock approximation 
are more generically associated with correlated quantum
fluctuations in the electronic configuration.  Nevertheless, we
believe that the approximation for the energy of the double-layer 
system which is obtained in the Hartree-Fock approximation is 
meaningful and that our results will be helpful in the interpretation
of double-layer capacitance studies.

We will assume that the electronic spins are fully polarized by 
the magnetic field and adopt a useful pseudospin language\cite{dlprl}
to describe the layer degree of freedom.  In this formalism the total
Hilbert space ${\cal H}$ is expressed as a direct product of the
orbital Hilbert space 
${\cal H}_o$ and
pseudospin Hilbert space ${\cal H}_s$. 
Choosing $\phi_{0,m}(x,y)$, symmetric-gauge two-dimensional 
free particle eigenstates\cite{lhreview} in the lowest Landau level, 
as a basis of ${\cal H}_o$ and assuming zero layer thickness, the
basis vectors of ${\cal H}$ can be written as
\begin{eqnarray}
\psi_{A,m} &=& \delta(z-z_A)\phi_{0,m}(x,y) ,
\label{f}
\end{eqnarray}
where $z_L=0$ and $z_R=d$.
Then the ${\cal H}_s$ spinors 
\begin{equation}
\label{a1} \left( \begin{array}{c} 1\\ 0\end{array}
\right)\; , \left( \begin{array}{c} 0\\ 1\end{array} \right)
\end{equation}
describe states in which an electron is localized 
in the left or right quantum well respectively.  This language suggests a 
magnetic analogy for the double-layer system. 
For example, with the definition in Eq.~(\ref{a1}) the $\hat{z}$-component
of the total pseudospin operator $\vec S$ is proportional to the 
difference in density between the layers since $<S^z>=(N_L-N_R)/2$.
The $\hat x$ and $\hat y$ components of the pseudospin operators
correspond to components of the density operator which are off diagonal
in layer indices; non-zero expectation values are possible
only when there is inter-layer phase coherence.
In a special limit of $d=0$ interactions between electrons in the 
same layer are identical to those between electrons in different layers 
and the Hamiltonian has $SU(2)$ symmetry: 
$[H,S^{\mu}]=0$ and eigenstates occur in multiplets with 
pseudospin quantum number $S$ and degeneracy $2S+1$.  For 
finite layer separation only $S_z$ is a good quantum number.

In the limit of large $d$, the equilibrium charge distribution is determined 
solely by electrostatic considerations and the functions (\ref{f})
describe  eigenstates of the
corresponding Hartree Hamiltonian. In this limit, 
it follows from Eq.~(\ref{eq:amd3})
that  the two-layers can be brought into equilibrium only if $N_R= N_0$,
{\em i.e.,} only if the electric field between the layers is
equal to zero.  
Any change in the gate voltage will result in a change in charge density
exclusively in the left well.  In the strong magnetic field limit 
considered here, equilibrium can be established only if the total
filling factors $\nu\equiv 2\pi\ell^2 N\in (\nu_0,1+\nu_0)$, 
where the threshold filling factor 
\begin{equation}
\nu_0=\frac{2\pi l^2 \epsilon}{|e|}E_0  
\label{c}
\end{equation}
and the magnetic length $\ell \equiv (\hbar / |e| B)^{1/2}$. Outside this
interval the left layer lowest Landau level is either empty or is
completely filled. Therefore,
\begin{equation}
\nu_R = \left\{ \begin{array}{ll} \nu &
\mbox{for $ \nu<\nu_0$}\\ \nu_0 & \mbox{for $ \nu_0\leq\nu\leq 1+\nu_0
$}\\ \nu -1 & \mbox{for $\nu>1+\nu_0 $} \end{array} \right.  .
\label{d}
\end{equation}
For smaller $d$ the charge distribution depends on intra and inter-layer 
correlations.  In what follows we use dimensionless units 
expressing energy in units of $e^2/4 \pi\epsilon \ell $
and lengths in units of the magnetic length, $\ell$.  
Deriving the Hartree-Fock self-consistent equation we will, for
simplicity, neglect 
tunneling between the two layers.  We will return to 
a discussion of the influence of tunneling later.
We will assume that the translational symmetry within each
two-dimensional layer is not broken so that the orbital 
degeneracy of the Landau levels is maintained.  

In our Hartree-Fock calculations we do not require $S_z$ to
be a good quantum number.  Allowing this  
symmetry to be broken gives rise to a much better variational
estimate of the ground state energy and results in states 
with spontaneous phase coherence between the layers.
We seek eigenstates $\left|\Psi\right>$ 
of the Hartree-Fock Hamiltonian with, generally, non-zero
expectation value of the $\hat{x}$ and $\hat{y}$ components of the pseudospin
operator.  The general form of the two orthogonal pseudospinors 
for the lower (`-') and higher (`+') energy
Landau levels are:
\begin{equation}
\alpha_-\equiv
\left( \begin{array}{c} \left<\psi_L|\Psi_-\right> \\ 
\left<\psi_R|\Psi_-\right>
\end{array} \right) = 
\left( \begin{array}{c} \cos\frac{\theta}{2} \\ {\rm
e}^{i\varphi}\sin\frac{\theta}{2}\end{array} \right) 
\label{g}
\end{equation}
and
\begin{equation}
\alpha_+\equiv
\left( \begin{array}{c} \left<\psi_L|\Psi_+\right> \\ 
\left<\psi_R|\Psi_+\right>
\end{array} \right) = 
\left( \begin{array}{c} \sin\frac{\theta}{2} \\ -{\rm
e}^{i\varphi}\cos\frac{\theta}{2}\end{array} \right).
\label{gg}
\end{equation}
The Hamiltonian in the pseudospin Hilbert space has a 
2$\times$2  matrix representation
\begin{equation}
H=\left( \begin{array}{cc} \varepsilon_{L}
& 0 \\ 0 & \varepsilon_{R} \end{array} \right) + \left(
\begin{array}{cc} \Sigma_{LL} & \Sigma_{LR} \\ \Sigma_{RL} &
\Sigma_{RR} \end{array} \right) , 
\label{h}
\end{equation}
where the Hartree potential appears in $\varepsilon_L$ and 
$\varepsilon_R$ and $\Sigma_{ij}$ are matrix elements of the exchange
self-energy. The orbital indices are omitted in Eq.~(\ref{g}-\ref{h}) since
the exchange self-energy is independent of the orbital quantum number 
of the Landau level, as 
we will explicitly prove. The self-consistent pseudospinor 
orientations, and consequently the charge distribution, 
can be determined by solving the Hartree-Fock equations 
iteratively using the expression for the self-energy given below.

\subsection{Total filling factor $\nu\le 1$}

In case when the total filling factor $\nu \le 1$ 
only the lower energy pseudospinor (\ref{g}) is occupied and we obtain
for the filling factor in the right layer
\begin{equation}
\nu_R=\nu\sin^2\frac{\theta}{2} .
\label{i}
\end{equation}
Including the Hartree self-energy and choosing the zero of energy so
that $\varepsilon_R = 0$
it follows directly from the Poisson that
\begin{equation}
\varepsilon_{L}=2d\left(\nu_0-\nu\sin^2\frac{\theta}{2}\right) .
\label{j}
\end{equation}
The evaluation of the exchange self-energy is more cumbersome;
we describe the derivation in detail for $\Sigma_{LL}$.  The same procedure
can be directly applied to all other self-energy matrix elements.  
Assuming that for the low-energy
pseudospinor all orbital states are occupied with probability $\nu$ 
we find that 
\begin{eqnarray}
\Sigma_{LL}(m,m')
&=&-\nu\sum_{n}\int {\rm d}\vec r \psi_{L,m}^*(\vec
r_1)\Psi_{-,n}^*(\vec r_2) \times \nonumber \\ &{}& \nonumber \\
&\times&\Psi_{-,n}(\vec r_1)\psi_{L,m'}(\vec r_2) V(\vec r_1-\vec r_2).
\label{k}
\end{eqnarray}
The fractional occupation results
from taking the zero temperature limit of a finite-temperature Hartree-Fock
expressions and occurs because of the Landau level degeneracy of the 
Hartree-Fock eigenvalues.  
Using (\ref{f}), (\ref{g}) and performing a Fourier transformation
of the Coulomb potential $V(\vec r_1-\vec r_2)$, Eq.~(\ref{k}) can be
rewritten as
\begin{eqnarray}
\Sigma_{LL}(m,m')&=&-\nu\cos^2\frac{\theta}{2} \sum_{n}\int
{\rm d}^2 r_{\perp} \int\frac{{\rm d}^2 q_{\perp}}{(2\pi)^2}
V_{eff}(\vec q_{\perp}) \times \nonumber \\ &{}& \nonumber \\ &\times&
\phi_{0,m}^*(\vec r_{\perp 1}){\rm e}^ {i\vec q_{\perp}\vec r_{\perp
1}}\phi_{0,n}(\vec r_{\perp 1}) \phi_{0,n}^*(\vec r_{\perp 2}){\rm e}^
{-i\vec q_{\perp}\vec r_{\perp 2}}\phi_{0,m'}(\vec r_{\perp 2}),
\label{l}
\end{eqnarray}
where
\begin{eqnarray} 
V_{eff}(\vec q_{\perp})
&=&\frac{e^2}{\epsilon}\int {\rm d}z \int\frac{d q_z}{2\pi}
\delta(z_1)\delta(z_2)\frac{{\rm e}^{iq_z(z_1-z_2)}}
{q_{\perp}^2+q_z^2}\nonumber \\ &=& \frac{e^2}{2\epsilon
|q_{\perp}|}.
\label{m}
\end{eqnarray}
The sum over $n$ in Eq.~(\ref{l}) can be evaluated analytically 
as shown in the Appendix, and is proportional to $\delta_{m',m}$.
Thus, the exchange self-energy is diagonal and independent of $m$ 
and every state in the Landau level has the same spinor as anticipated.
Finally we obtain
\begin{equation}
\Sigma_{LL}=-\nu\cos^2\frac{\theta}{2}I_A ,
\label{q}
\end{equation}
where for the case of Coulomb interactions the 
intra-layer exchange integral $I_A=\sqrt{\pi/2}$.

A similar calculation shows that $\Sigma_{RR}$ is given 
by the same expression with  $\cos^2\theta/2$ replaced by $\sin^2\theta/2$.
For the inter-layer exchange self-energies , the potential $V_{eff}$ is
modified because of the layer separation $d$.
For Coulomb interactions the inter-layer exchange integral is 

\begin{equation}
I_E=\int_0^\infty {\rm d}q
\exp\left(-\frac{q^2}{2}-dq\right).
\label{r}
\end{equation}

Using the explicit expressions for the Hartree and
exchange self-energies derived above in Eq.~(\ref{h}) we obtain
the  Hamiltonian 
\begin{eqnarray}
&&H = \left( \begin{array}{cc}
2d\left(\nu_0-\nu\sin^2\frac{\theta}{2}\right) & 0 \\ & \\ 0 & 0
\end{array} \right) - \nonumber \\ & & \nonumber \\ & & \nonumber \\
&-& \nu \left( \begin{array}{cc} \cos^2\frac{\theta}{2}I_A &
\sin\frac{\theta}{2}\cos\frac{\theta}{2} {\rm e}^{i\varphi}I_E \\ & \\
\sin\frac{\theta}{2}\cos\frac{\theta}{2}{\rm e}^{-i\varphi}I_E &
\sin^2\frac{\theta}{2}I_A \end{array} \right) .
\label{s}
\end{eqnarray}
The eigenfunctions of this Hamiltonian are easily found
by expanding it in terms of Pauli spin matrices:
\begin{equation}
H=H_0+\vec{{\cal B}}\vec{\sigma},
\label{t}
\end{equation}
where
\begin{equation}
H_0=\frac{\varepsilon_L}{2}-\frac{\nu}{2} I_A
\label{u}
\end{equation}
and the effective `Zeeman' field
$\vec{\cal B}$ has components
\begin{eqnarray}
{\cal B}_x&=&-\frac{\nu}{2}\sin\theta\cos\varphi I_E \nonumber\\ 
{\cal B}_y&=&-\frac{\nu}{2}\sin\theta\sin\varphi I_E \nonumber\\
{\cal B}_z&=&\frac{\varepsilon_L}{2}-\frac{\nu}{2}\cos\theta I_A .
\label{v}
\end{eqnarray}
The low energy eigenspinor of $H$ will be the spinor
which is aligned with $\vec{\cal B}$.   Self-consistency is 
therefore achieved when $\vec{\cal B}$ has the same 
orientation as the spinor from which the exchange self-energy 
was constructed. This condition reduces to an algebraic equation for
the polar angle $\theta$:
\begin{equation}
\tan\theta =\frac{\nu\sin\theta I_E}{\nu\cos\theta I_A - \epsilon_L}. 
\label{w}
\end{equation}
If $\theta\neq0,\pi$, exchange electron-electron interactions lead
to phase coherence between electrons in
different layers.  The direction of the ground state 
pseudospin is specified by the angles $\theta$
and  $\varphi$.  Note that the azimuthal angle $\varphi$ is arbitrary.

\subsection{Total filling factor $\nu>1$}

At $\nu>1$ all states in the low energy Landau level are full
and the high energy Landau level is partially occupied.
The contribution of the higher energy Landau level to both 
Hartree and exchange self-energies has to be included.
For example, the filling factor in the right layer for $\nu > 1$ is 
given by 
\begin{equation}
\nu_R=\sin^2\frac{\theta}{2}+(\nu-1)\cos^2\frac{\theta}{2}.
\label{x}
\end{equation}
We again obtain degenerate Landau levels.  In this case we 
find that the pseudospinor Hamiltonian is given by
\begin{eqnarray}
H &=& \left( \begin{array}{cc}
2d\left(\nu_0-\sin^2\frac{\theta}{2}-(\nu-1)\cos^2\frac{\theta}{2}\right)
& 0 \\ & \\ 0 & 0 \end{array} \right) - \nonumber \\ & & \nonumber \\
& & \nonumber \\ &-& \left( \begin{array}{cc}
\cos^2\frac{\theta}{2}I_A & \sin\frac{\theta}{2}\cos\frac{\theta}{2}
{\rm e}^{i\varphi}I_E \\ & \\
\sin\frac{\theta}{2}\cos\frac{\theta}{2}{\rm e}^{-i\varphi}I_E &
\sin^2\frac{\theta}{2}I_A \end{array} \right) - \nonumber \\ & &
\nonumber \\ & & \nonumber \\ -(\nu-1&)& \left( \begin{array}{cc}
\sin^2\frac{\theta}{2}I_A & -\sin\frac{\theta}{2}\cos\frac{\theta}{2}
{\rm e}^{i\varphi}I_E \\ & \\
-\sin\frac{\theta}{2}\cos\frac{\theta}{2}{\rm e}^{-i\varphi}I_E &
\cos^2\frac{\theta}{2}I_A \end{array} \right).
\label{y}
\end{eqnarray}
When this is expanded in terms of Pauli spin matrices it 
results in a effective Zeeman field given by 
\begin{eqnarray}
{\cal
B}_x&=&-\frac{2-\nu}{2}\sin\theta\cos\varphi I_E \nonumber\\ {\cal
B}_y&=&-\frac{2-\nu}{2}\sin\theta\sin\varphi I_E \nonumber\\ {\cal
B}_z&=&\frac{\varepsilon_L}{2}-\frac{2-\nu}{2}\cos\theta I_A .
\label{z}
\end{eqnarray}

\section{Unrestricted Hartree Fock Approximation: Total Energy}

Eq.~(\ref{w}) often has more than one solution.   The best 
unrestricted Hartree-Fock approximation to the ground state of the double-layer 
system is the solution with the lowest energy.  In the Hartree-Fock
approximation the total energy $E_{TOT}$ for two-dimensional electron 
systems in the strong magnetic field limit can be separated into
electrostatic (Hartree) and exchange contributions.  
(The quantized kinetic energy is absorbed into the 
zero of energy and correlation effects are neglected in the 
Hartree-Fock approximation.)  For a given $\nu$ 
constant the Hartree energy is (up to an arbitrary constant) proportional to
the energy density in the intra-layer electric field:
\begin{equation}
E_{H} = \frac{\epsilon dAE_1^2}{2}.
\label{A}
\end{equation}
The electric field $E_1$ can be expressed as a function of
pseudospin orientation using Eqs.~(\ref{a}),(\ref{c}),(\ref{i}),
and (\ref{x}).  Using the dimensionless variables introduced in section II,
\begin{equation}
\frac{E_{H}}{A} = \left\{\begin{array}{ll}
\frac{d\left(\nu_0-\nu\sin\frac{\theta}{2}\right)^2}{2\pi} & \mbox{for
$ \nu\leq 1$}\\ 
 & \\
\frac{d\left(\nu_0-\sin^2\frac{\theta}{2}-(\nu-1)
\cos^2\frac{\theta}{2}\right)^2}{2\pi}
& \mbox{for $ \nu>1$} \end{array}\right.  .
\label{B}
\end{equation}
In evaluating the exchange energy it is necessary to avoid double-counting
electron-electron interactions.  For $\nu \le 1$ only the low-energy
pseudospinor is occupied while for $\nu > 1$ both spinors are
occupied and we find that 
\begin{equation}
\frac{E_{X}}{NA} = \left\{
\begin{array}{ll} \frac{1}{2}\alpha_-^{\dagger}H_X\alpha_- & \mbox{for $
\nu\leq 1$}\\ & \\ \frac{1}{2\nu}\left(\alpha_-^{\dagger}H_X\alpha_-
+(\nu-1) \alpha_+^{\dagger}H_X\alpha_+\right)& \mbox{for $ \nu>1$}
\end{array} \right. ,
\label{D}
\end{equation}
where $H_X$ is the exchange contribution to the Hartree-Fock
Hamiltonian.  (Explicit expressions
for $H_X$ were derived for both $\nu\leq 1$ and
$\nu>1$ in the previous section.) 
Using Eqs.~(\ref{s}),(\ref{y}),(\ref{D}) and the definition of the
filling factor we obtain
the following  results, in dimensionless units,
for the dependence of the exchange energy 
on pseudospin orientation.  For $\nu\leq 1$
\begin{equation}
\frac{E_{X}}{A} = -\frac{\nu^2}{4\pi}\left(I_A(\sin^4\frac{
\theta}{2}+\cos^4\frac{\theta}{2})+2I_E\sin^2\frac{\theta}{2}
\cos^2\frac{\theta}{2}\right)
\label{E}
\end{equation}
and for $\nu>1$
\begin{eqnarray}
\frac{E_{X}}{A} &=& -\frac{1}{4\pi}\left(I_A(\sin^4\frac{
\theta}{2}+\cos^4\frac{\theta}{2})\left(1+(\nu-1)^2\right)+\right.
\nonumber \\
 &{}&\nonumber \\
  &+& \left.  2\sin^2\frac{\theta}{2}
\cos^2\frac{\theta}{2}\left(I_A(\nu-1)+I_E(2-\nu)^2 \right) \right) .  
\label{F}
\end{eqnarray}
Note, that minimizing the total energy with respect to the angle
$\theta$, {\em i.e.}, solving the equation
\begin{equation}
\frac{{\rm d}E_{TOT}/A}{{\rm d}\theta}= \frac{{\rm d}(E_{H}+
E_{X})/A}{{\rm d}\theta}=0 ,
\label{G}
\end{equation}
yields an equation identical to that resulting from requiring the
pseudospinor to self-consistently solve 
Eq.~(\ref{w}).  If more than one solution occurs we choose the solution
with the lowest energy.

\section{Numerical Results}

We find that solutions to Eq.~(\ref{G}) can occur at $\theta = 0$,
at $\theta = \pi$ and at most at one $\theta \in (0,\pi)$.
$\theta=0$ solutions correspond, for $\nu < 1$, to all the electrons being in the 
left well, while $\theta=\pi$ solutions correspond, for $\nu < 1$,
to all electrons being in the right well.  For $\nu > 1 $ these two 
solutions correspond to the full Landau levels in the 
left and right well  respectively.  For $\theta \in (0,\pi)$ both layers
are partially occupied and in equilibrium.  These solutions occur 
when 
\begin{equation}
\cos\theta=\left\{ \begin{array}{ll}
\frac{d(2\nu_0-\nu)}{\nu(I_A-I_E-d)} & \mbox{for $ \nu\leq 1$}\\ & \\
\frac{d(2\nu_0-\nu)}{(2-\nu)(I_A-I_E-d)} & \mbox{for $ \nu>1$}
\end{array} \right.  .
\label{H}
\end{equation}
From Eq.~(\ref{H}) we see that both layers can be partially occupied 
only in the region of the $\nu - d$ plane where
the absolute value of the right-hand side of Eq.~(\ref{H}) is less
than 1.  (Recall that $I_E$ has a dependence on $d$ which is
implicit in these equations.)  
For $0\le\nu\le 1$ the boundary of this region is defined by the curves
\begin{equation}
d=\left\{ \begin{array}{ll}
\frac{\nu}{\nu-\nu_0}\frac{I_A-I_E}{2} & \mbox{for $ \nu\leq
2\nu_0$}\\ & \\ \frac{\nu}{\nu_0}\frac{I_A-I_E}{2} & \mbox{for $
\nu>2\nu_0$} \end{array} \right.  ,
\label{I}
\end{equation}
while for $1\le\nu\le 2$ the boundary is defined by the 
curves
\begin{equation}
d=\left\{ \begin{array}{ll}
\frac{2-\nu}{1-\nu_0}\frac{I_A-I_E}{2} & \mbox{for $ \nu\leq
2\nu_0$}\\ & \\ \frac{2-\nu}{1-\nu+\nu_0}\frac{I_A-I_E}{2} & \mbox{for
$ \nu>2\nu_0$} \end{array} \right.  .
\label{J}
\end{equation}
The solution with the two layers in equilibrium is always 
lowest in energy whenever it is self-consistent, {\em i.e.} whenever
a local energy minimum occurs for $\theta \in (0 , \pi)$.
When this solution doesn't exist, the polar angle $\theta = 0$ 
for $\nu>2\nu_0$ and $\theta=\pi$ for $\nu<2\nu_0$.
In the cases $\nu_0\le 0$ or $\nu_0\ge 1$\, $\theta = 0$ or $\theta = \pi$
throughout the strong magnetic field regime.  
In Figure~\ref{f14}, Figure~\ref{f12} and Figure~\ref{f34} 
we show results obtained at 
$\nu_0 =1/4$, $\nu_0 = 1/2$ and $\nu_0 = 3/4$ when there 
is no inter-layer hopping.
The upper panel of each figure is a phase diagram which shows
the state of the system as a function of layer separation and total
filling factor.
Note that there
is a mirror symmetry along the line $\nu=1$ between the phase diagrams
for $\nu_0=x$ ($x<1/2$) and $\nu_0=1-x$.
In Region I in these phase diagrams the two layers are not
in equilibrium.  In the left Region I all the electrons are 
in the right layer and the Hartree-Fock eigenenergy for the 
left layer lies above the chemical potential.  In the right 
Region I the left Landau level is completely filled and its 
Hartree-Fock eigenenergy lies below the Fermi energy.
In Region II, $\theta \in (0, \pi)$, and each Hartree-Fock
eigenfunction is a coherent linear combination of states localized in the 
two wells.  We do not believe that this spontaneous phase coherence 
exists throughout the the entire Region II as indicated schematically
by the dashed lines in Region II.  For example, for the case
$\nu_0 = 1/2$, $\nu =1$, which has been studied extensively both 
theoretically\cite{dlprl,zee,ezawa} and
experimentally,\cite{murphy,mansour} spontaneous coherence is expected to
occur only for $d < \approx 2$.   It is very difficult to predict
theoretically where, within Region II, spontaneous phase coherence will
occur; the dashed lines in the  figures are intended to suggest only that
it is most likely near 
$\nu =1$ and at small layer separations
where the Hartree-Fock approximation is most reliable.
We believe that this question is best addressed experimentally.
Stimulating such experiments is part of the motivation for 
this work.

The middle panels in Figure~\ref{f14}, Figure~\ref{f12}, and Figure~\ref{f34}
shows the optimal (self-consistent) filling factor in the 
right well as a function of the total filling factor for 
$d=1$, $d=5$, and $d \to \infty$.  
These three layer separations correspond to strongly 
coupled layers, weakly coupled layers and decoupled layers.
For $d \to \infty$ all the charge goes into the right
layer until the electric field reaches zero between the layers.
When this point is reached all the incremental charge goes to the left layer
until its Landau level is filled.  Only then does the filling of the 
right layer resume.  Exchange  
tends to favor unequal layer occupations
except at the point where the layers are balanced, $\nu = 2 \nu_0$,  
so that the left-layer is not occupied until larger total filling factors
at smaller $d$.  Once the occupation of the left layer begins,
the right layer occupation gradually {\em decreases} as the left layer 
Landau level is filled.  

The bottom panels of Figure~\ref{f14} , Figure~\ref{f12}, and
Figure~\ref{f34}  show the dependence of the 
Eisenstein ratio $R_E$ on total filling 
factor. In Region II both layers are in equilibrium 
and Eqs.~(\ref{eq:amd5}-\ref{eq:amd7})
apply. From Eqs.~(\ref{E}) and (\ref{F}) we obtain the Hartree-Fock values
for the length parameters
\begin{eqnarray}
d_{LL} &=& d_{RR} = -\frac{I_A}{2}\nonumber\\
d_{LR} &=& d_{RL} = -\frac{I_E}{2} 
\label{eisl}
\end{eqnarray}
and the Hartree-Fock Eisenstein ratio reads
\begin{equation}
R_E=-\frac{I_A-I_E}{2(d-I_A+I_E)}.
\label{eisr}
\end{equation}
For large layer separations $d\gg I_A$ 
and inter-layer coupling can be neglected so that $R_E$ 
is proportional to the reciprocal value to the compressibility
of an individual 2D layer.  For the Coulomb interaction
the Hartree-Fock theory in this limit 
gives $R_E = - I_A/2d = - \sqrt{\pi/8}/d $, missing the
anomalies\cite{jimcompress} 
associated with incompressible fractional quantum Hall states seen 
experimentally.  At smaller $d$ the electrostatic term in
the denominator becomes less dominant and inter-layer interactions
become important.  For small $d$, \, $I_A - I_E = d - d^2 \sqrt{\pi/8} 
+ \cdots $ so that in this limit the Hartree-Fock theory gives 
$R_E = - \sqrt{2/\pi}/d$, diverging for $ d \to 0$.
The Hartree-Fock Eisenstein ratio within Region II
is a negative monotonically increasing 
function of $d$ for all $d\in(0,\infty)$, as shown in Figure~\ref{fre}.
Neglecting the inter-layer interactions yields an unphysical divergence
of $R_E$ at $d=I_A$.

In the discussion of equilibrium properties of
the double-layer electron system presented above,
tunneling between the 2D layers was neglected.
In a tight-binding model, the tunneling contribution to the 
Hartree-Fock Hamiltonian is 
\begin{eqnarray}
&&H_t = \left( \begin{array}{cc} 0 & t \\ t
& 0 \end{array} \right)
\label{N}
\end{eqnarray}
where $t$ is a phenomenological parameter which is in practise
chosen to match either experimental or calculated values
of the splitting between the two-lowest subbands of the double-layer
system.  The self-consistent procedure derived 
for $t=0$, is readily generalized to include this term
in the Hamiltonian.  We find that for $t \ne 0$ 
both layers are partially filled and in equilibrium
throughout the strong magnetic field regime.  The tunneling term
in the Hamiltonian favors equal layer densities 
and therefore competes 
with the exchange electron-electron interactions. The filling factor
$\nu_R$ and the Eisenstein ratio $R_E$ as a function of $\nu$
are shown in Figure~\ref{ft} for several values of $t$ and for $\nu_0=1/2$.
Note that the steps in $R_E$ associated with establishing equilibrium
between the two layers are smeared by tunneling.

\section{Conclusions}

In this article we have shown how electron-electron interactions
beyond a simple electrostatic approximation 
influence the dependence on a remote gate voltage of 
the partitioning of electric charge in a double-layer 
system.  Our calculations are based on an unrestricted Hartree-Fock
approximation which can introduce inter-layer correlations by
forming a broken symmetry state with spontaneous inter-layer 
phase coherence.   We have made contact with potential experiments 
by expressing our results in terms of the Eisenstein ratio, which
is proportional to the rate of charge transfer between layers 
when the gate voltage is varied.  Our calculations 
demonstrates the essential role of inter-layer correlations;
if they were neglected in our calculations
the Eisenstein ratio would have an unphysical
divergence at $d=\sqrt{\pi/2}$.  The Hartree-Fock 
approximation we use has deficiencies that are known to be important 
in this system.  In particular, it does not capture the anomalies
in the Eisenstein ratio which are associated with the 
fractional quantum Hall effect.  However, we believe that our 
calculation provides a useful qualitative picture
which will be helpful in guiding and interpreting experimental studies
of coupled double-layer electron systems.   

\acknowledgments 
We acknowledge helpful interactions with J.P. Eisenstein.
This work has been supported by the National
Science Foundation through grants NSF INT-9106888 and 
DMR-9416906.  

\section{Appendix} The sum over $n$ in Eq.~(\ref{l}) can be
calculated using known identities for symmetric-gauge eigenfunctions.
It is useful to introduce a factor $G_{i,j}(k)$ ($G$ is a function of
complex wavevector $k=k_x+ik_y$) defined as
\begin{equation}
G_{i,j}(k) =
\left(\frac{j!}{i!}\right)^{1/2}\left(\frac{-ik}{\sqrt{2}}\right)^{i-j}
L_j^{i-j}\left(\frac{k\overline{k}}{2}\right) ,
\label{n}
\end{equation}
where $L_j^{i-j}\left(\frac{k\overline{k}}{2}\right)$ is the
generalized Laguerre polynomial.  The relation between $G$ and matrix
elements of $\exp(i{\vec k}\cdot {\vec r})$ reads
\begin{equation}
\int d^2 r \phi_{0,i}^*(r){\rm e}^{i{\vec
k}\cdot {\vec r}}\phi_{0,j}(r)= {\rm e}^{\frac{-|k|^2}{2}}G_{i,j}(kl).
\label{o}
\end{equation}
Then, since 
\begin{equation}
\sum_n G_{i,n}(k_1)G_{n,j}(k_2)={\rm e}^{\frac{-\overline{k}_1k_2}{2}}
G_{i,j}(k_1+k_2)
\label{p}
\end{equation}
we obtain
\begin{equation}
\sum_{n} \phi_{0,m}^*(\vec r_{\perp
1}){\rm e}^ {i\vec q_{\perp}\vec r_{\perp 1}}\phi_{0,n}(\vec r_{\perp
1}) \phi_{0,n}^*(\vec r_{\perp 2}){\rm e}^ {-i\vec q_{\perp}\vec
r_{\perp 2}}\phi_{0,m'}(\vec r_{\perp 2})=
\delta_{m,m'}\exp(-q^2_{\perp}).
\label{pp}
\end{equation}

\begin{figure}
\caption{Simplified band diagram for a gated  double-quantum-well
structure in a strong perpendicular magnetic field. }
\label{fdw}
\end{figure}

\begin{figure}
\caption{Results for the threshold filling factor
$\nu_0=1/4$ (no inter-layer hopping):  
a) Hartree-Fock phase diagram. 
b) Filling factor of the right quantum well as a function
of the total filling factor for
$d=1, 5, \infty$.  c) Eisenstein ratio as a function
of the total filling factor for the same
layer separations as in b).}
\label{f14}
\end{figure}

\begin{figure}
\caption{Results for the threshold filling factor
$\nu_0=1/2$ (no inter-layer hopping):  
a) Hartree-Fock phase diagram. 
b) Filling factor of the right quantum well as a function
of the total filling factor for
$d=1, 5, \infty$.  c) Eisenstein ratio as a function
of the total filling factor for the same
layer separations as in b).}
\label{f12}
\end{figure}

\begin{figure}
\caption{Results for the threshold filling factor
$\nu_0=3/4$ (no inter-layer hopping):  
a) Hartree-Fock phase diagram. 
b) Filling factor of the right quantum well as a function
of the total filling factor for
$d=1, 5, \infty$.  c) Eisenstein ratio as a function
of the total filling factor for the same
layer separations as in b).}
\label{f34}
\end{figure}

\begin{figure}
\caption{Eisenstein ratio in Region II as a function of the layer 
separation with inter-layer interaction taken into account (solid line)
and for $I_E=0$ (dotted line).}
\label{fre}
\end{figure}

\begin{figure}
\caption{Results for the threshold filling factor
$\nu_0=1/2$, for layer separation $d=1$ and for selected values
of the inter-layer hopping parameter $t$:  
a) Filling factor of the right quantum well as a function of
the total filling factor.
b) Eisenstein ratio as a function of the total filling factor.}
\label{ft}
\end{figure}


\begin{references}

\bibitem{dragexp} T.J. Gramila, J.P. Eisenstein, A.H. MacDonald,
L.N. Pfeiffer, and K.W. West, Phys. Rev. Lett. {\bf 66}, 1216 (1991);
Phys. Rev. B {\bf 47}, 12957 (1993); U. Sivan, P.M. Solomon,
and H. Shtrikman, Phys. Rev. Lett. {\bf 68}, 1196 (1992).

\bibitem{dragth} P.J. Price, Physica B {\bf 117}, 750 (1983);
H.C. Tso, P. Vasilopoulos, and F.M. Peeters, Phys. Rev. Lett.
{\bf 68}, 2516 (1992); A.-P. Jauho, and H. Smith, Phys. Rev.
B {\bf 47}, 4420 (1993); L. Zheng, and A.H. MacDonald,
Phys. Rev. B {\bf 48}, 8203 (1993); K. Flensberg and Ben
Yu-Kuang Hu, Phys. Rev. Lett. {\bf 73}, 3572 (1994).

\bibitem{inplane} J.A. Simmons, S.K. Lyo, N.E. Harff, and J.F. Klem,
Phys. Rev. Lett. {\bf 73}, 2256 (1994); S.K. Lyo, Phys. Rev. B 
{\bf 50}, 4965 (1994);
G.~S. Boebinger, A.~Passner, L.~N. Pfeiffer, and K.~W. West,
Phys. Rev. B {\bf 43}, 12673 (1990); J. Hu and A.H. MacDonald,
Phys. Rev. B {\bf 46}, 12554 (1992); Y. Berk, A. Kamenv,
A. Palevksi, L.N. Pfeiffer, and K.W. West, Phys. Rev. B {\bf 51},
2604 (1995).

\bibitem{interplane} J.P. Eisenstein {\em et. al.}, Phys. Rev.
B {\bf 44}, 6511 (1991); J.A. Simmons {\em et. al.} Phys. Rev.
B {\bf 47}, 15741 (1993); S.K. Lyo and J.A. Simmons, J. Phys. Condens.
Matter {\bf 5}, L299 (1993).

\bibitem{dlprl} See for example
Kun Yang, K. Moon, L. Zheng, A.H. MacDonald, S.M. Girvin, D.
Yoshioka, and Shou-Cheng Zhang, Phys. Rev. Lett. {\bf 72}, 732, 1994 and
work cited therein.

\bibitem{murphy} S.Q. Murphy, J.P. Eisenstein, G.S. Boebinger,
L.N. Pfeiffer, and K.W. West, Phys. Rev. Lett. {\bf 72}, 728 (1994);
A.H. MacDonald, P.M. Platzman, and G.S. Boebinger,
Phys. Rev. Lett. {\bf 65}, 775 (1990);
G.S. Boebinger {\em et. al.}, Phys. Rev. Lett. {\bf 64}, 1793 (1990);
Y.W. Suen {\em et. al.}, Phys. Rev. B {\bf 44}, 5947 (1991).

\bibitem{jimcompress} J.P. Eisenstein, L.N. Pfeiffer, and K.W. West,
Phys. Rev. B {\bf 50}, 1760 (1994); J.P. Eisenstein, L.N. Pfeiffer, and
K.W. West, Phys. Rev. Lett. {\bf 68}, 674 (1992).

\bibitem{dlnum} Reliable calculations of the ground state energy
are possible using exact diagonalization for finite-size systems.
See for example Chakraborty and P. Pietil\"{a}inen,
Phys. Rev. Lett., {\bf 59}, 2784 (1987); E.H. Rezayi and F.D.M.
Haldane, Bull. Am. Phys. Soc. {\bf 32}, 892 (1987);
Song He, S. Das Sarma and X.C. Xie, Phys. Rev. B {\bf
47}, 4394 (1993); D. Yoshioka, A.H. MacDonald, and S.M. Girvin, Phys.
Rev. B {\bf 39}, 1932 (1989).

\bibitem{amdreview} For a brief review of the fractional quantum Hall
effect in double-layer systems see
A.H. MacDonald, Surface Science {\bf 229}, 1 (1990).

\bibitem{lian} L. Zheng and A.H. MacDonald, Phys. Rev. B
{\bf 49}, 5522 (1994) and related unpublished calculations. 

\bibitem{lhreview} See for example, A.H. MacDonald,
{\em Introduction to the Physics of the Quantum Hall Regime}
Proceedings of the 1994 Les Houches Summer School on Mesoscopic
Physics, to be published by North Holland.  Indiana University
Preprint IUCM-94014.

\bibitem{zee} X.G. Wen and A. Zee, Phys. Rev. Lett. {\bf
69}, 1811 (1992); X.G. Wen and A. Zee, Phys. Rev. B {\bf 47}, 2265
(1993).

\bibitem{ezawa} Z.F. Ezawa  and A. Iwazaki, Int. J. of
Mod. Phys. B,
{\bf 19}, 3205 (1992); Z.F. Ezawa and A. Iwazaki, Phys. Rev. B {\bf
47}, 7295 (1993); Z.F. Ezawa, A. Iwazaki, Phys. Rev. B {\bf 48}, 15189
(1993).

\bibitem{mansour} T.S. Lay, Y. W. Suen, H.C. Manoharan,
X. Ying, M.B. Santos, and M. Shayegan, Phys. Rev. B {\bf 50}, 17725
(1994). 

\end{references}
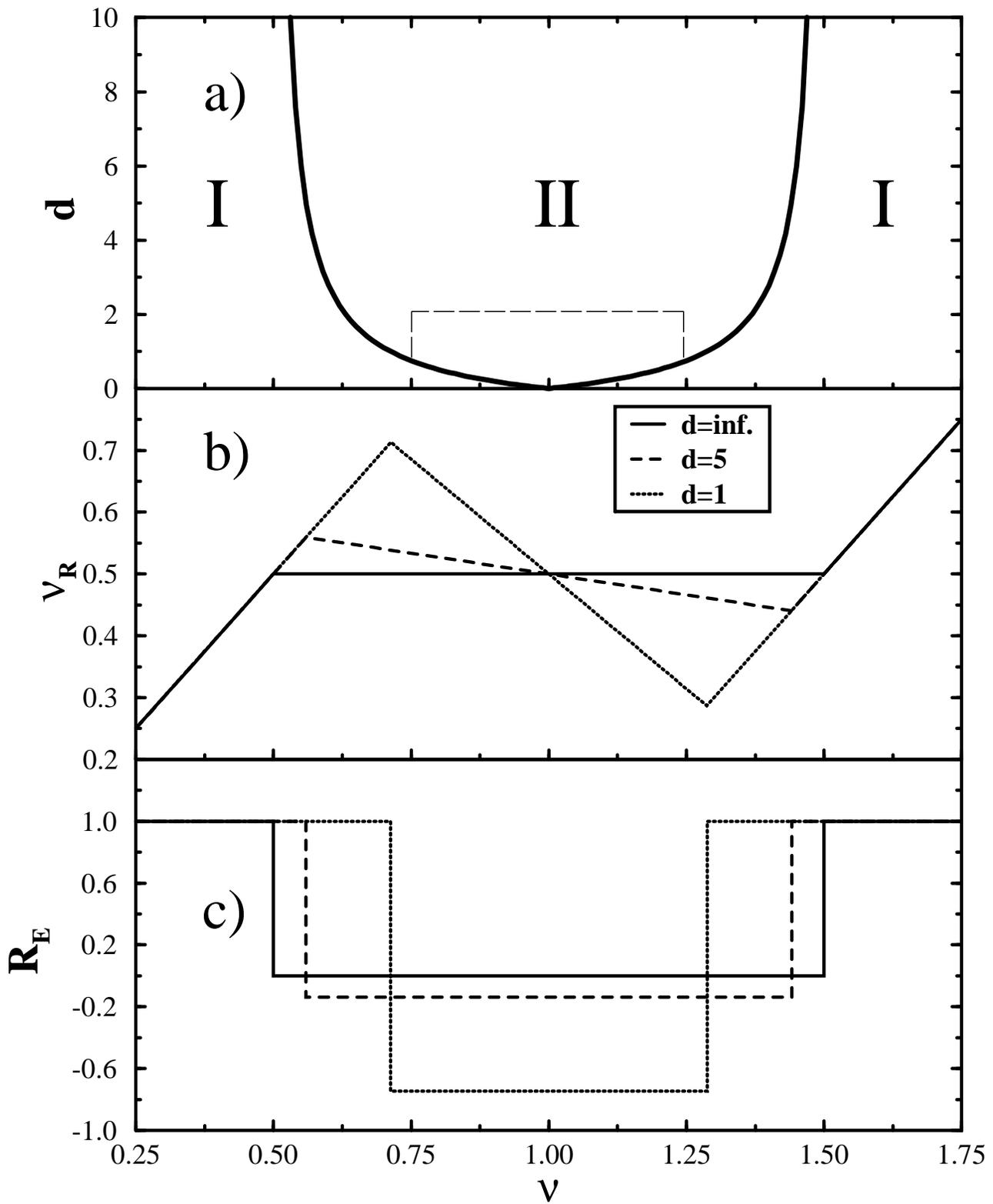
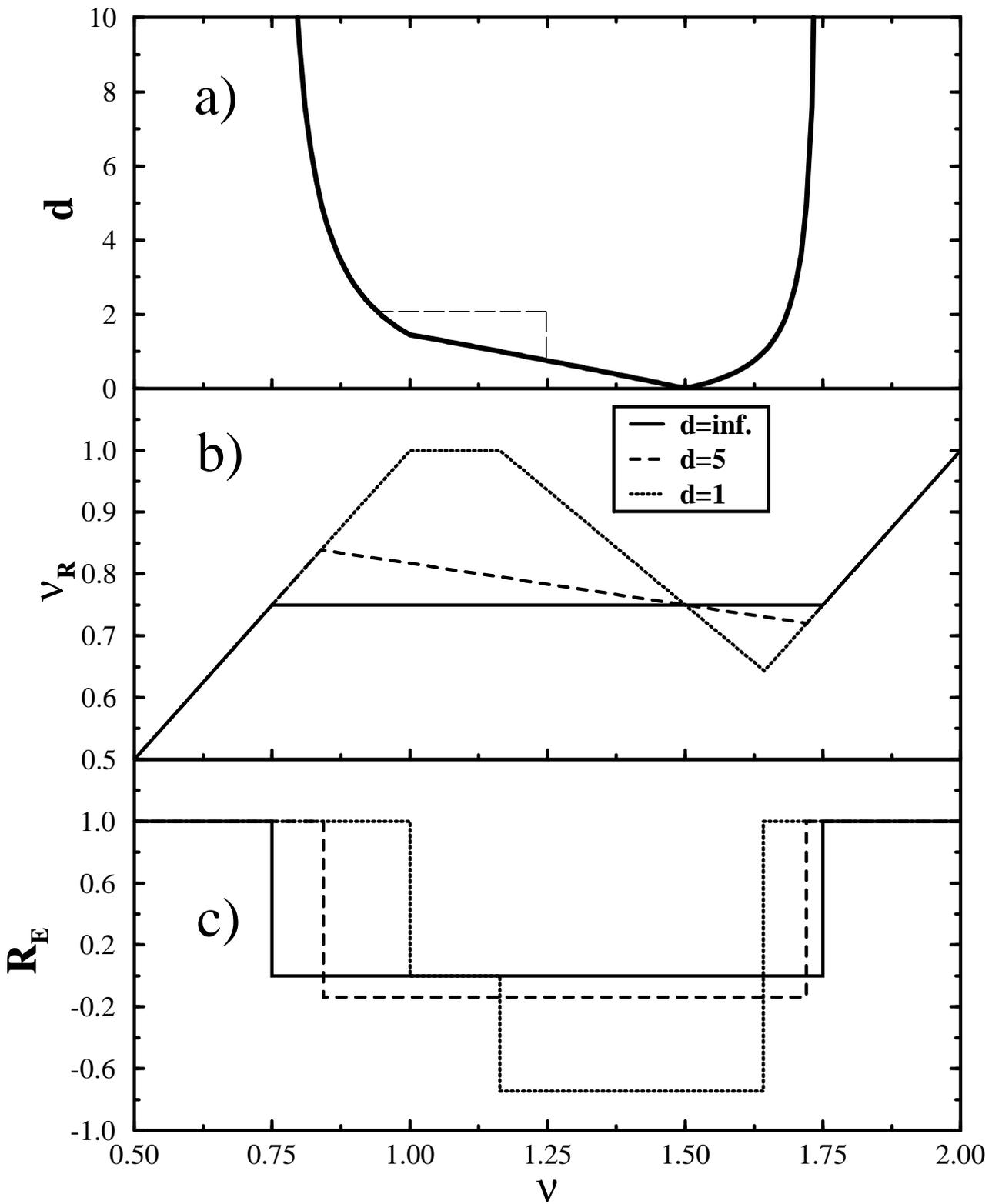
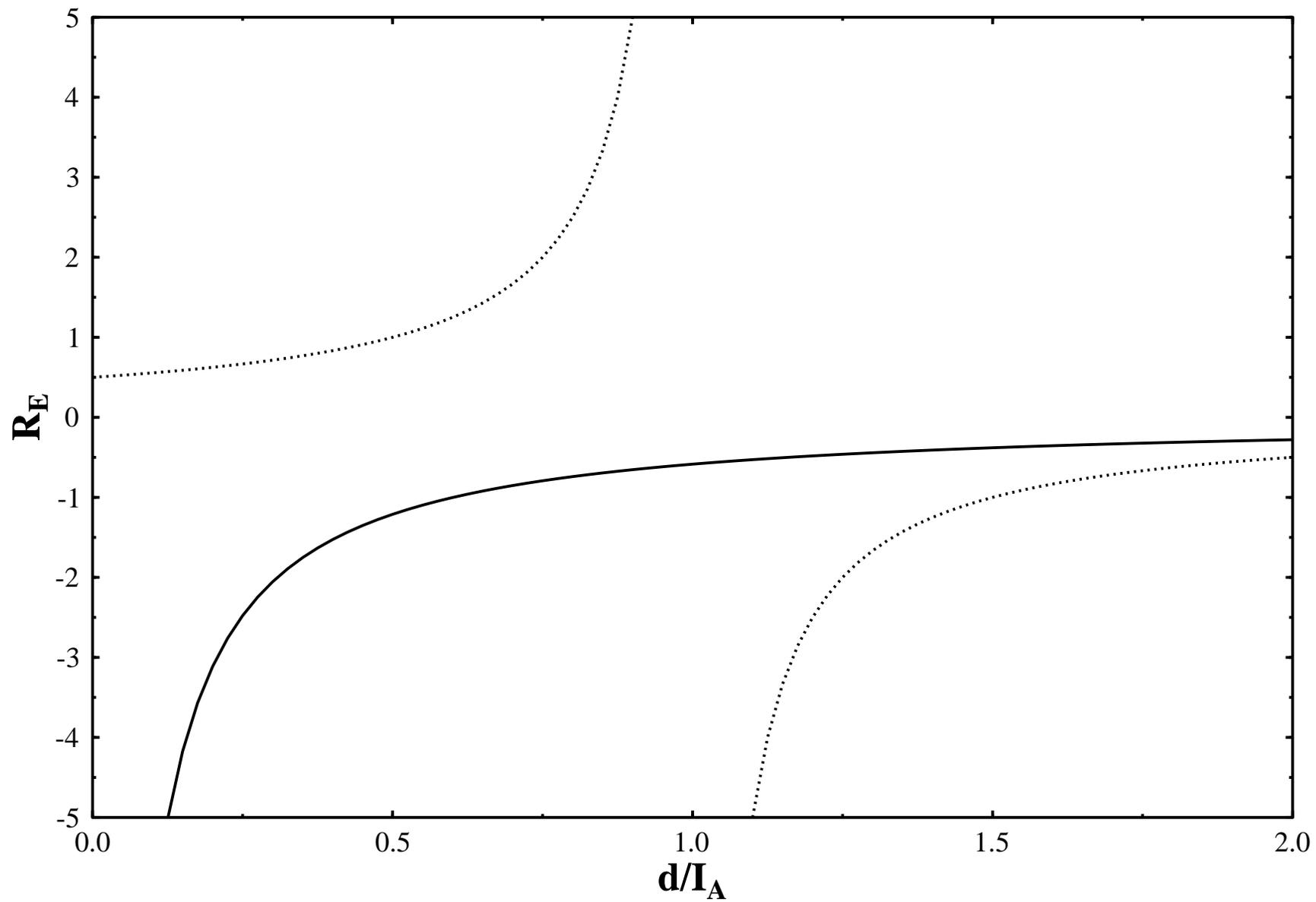
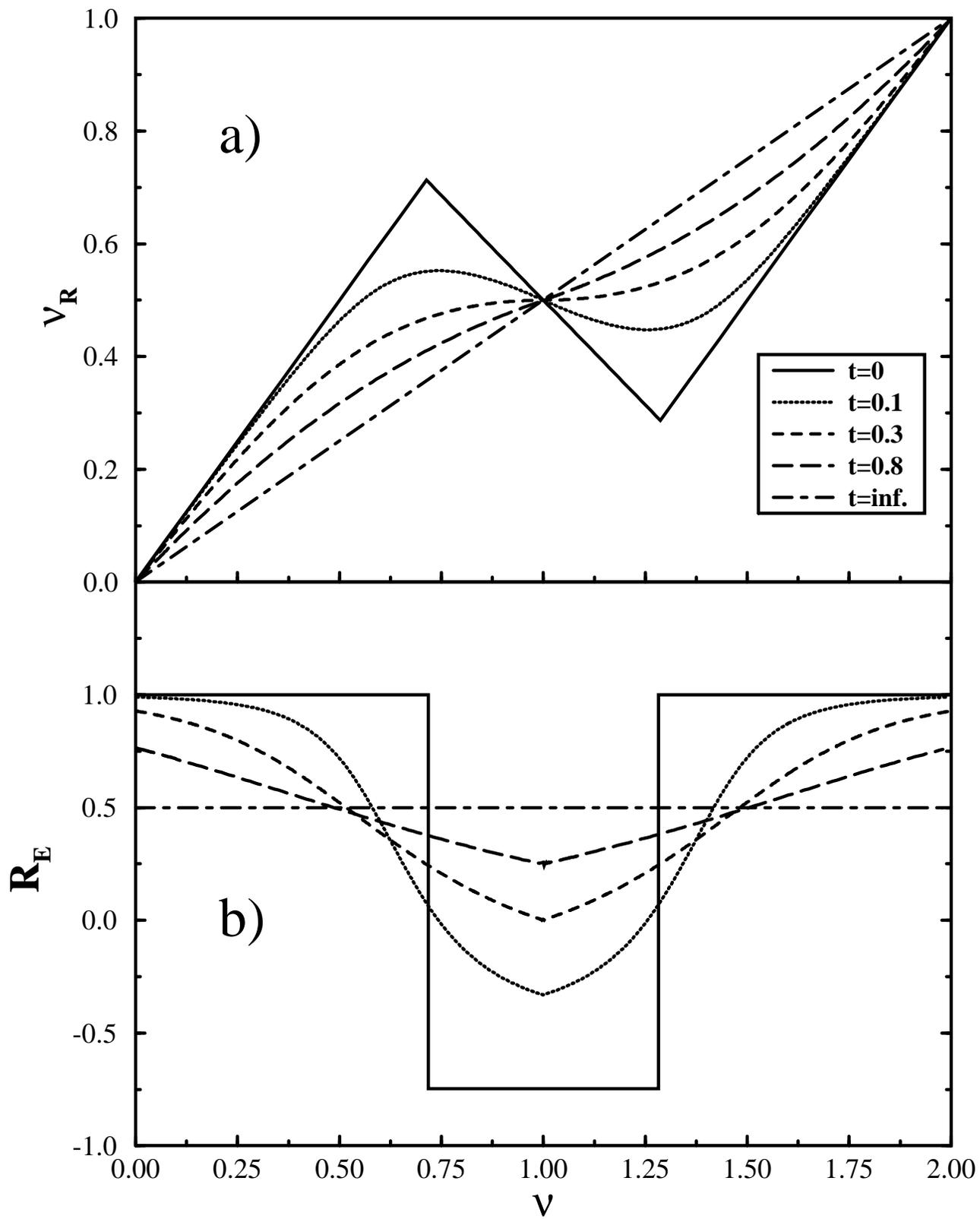
\end{document}